\documentclass[]{aa} 
\usepackage[]{natbib}
\bibpunct{(}{)}{;}{a}{}{,} 
\usepackage[varg]{txfonts}
\usepackage{graphicx}
\usepackage{color}
\usepackage{wasysym}
\usepackage{amsmath}
\usepackage{morefloats}
\usepackage{balance}
\usepackage{nameref}
\usepackage{amsmath}
\usepackage{hyperref}
\hypersetup{ 
    colorlinks=true,
    linkcolor=blue,
    filecolor=blue,      
    urlcolor=blue,
    citecolor=blue,
}

\newcommand\maps{{Meteorit. Planet. Sci. }}
\newcommand\epsl{{Earth Planet. Sci. Lett. }}

\begin{document}

\title{A water budget dichotomy of rocky protoplanets from \textsuperscript{26}Al-heating}
\titlerunning{A water budget dichotomy of rocky protoplanets from \textsuperscript{26}Al-heating} 
\authorrunning{Lichtenberg et al.}

\author{
Tim Lichtenberg,\inst{\ref{inst1},\thanks{Corresponding author, present address: Atmospheric, Oceanic and Planetary Physics, University of Oxford, Parks Rd, Oxford OX1 3PU, United Kingdom, \email{tim.lichtenberg@physics.ox.ac.uk}}}
Gregor J. Golabek,\inst{\ref{inst2}}
Remo Burn,\inst{\ref{inst3}}
Michael R. Meyer,\inst{\ref{inst4}}\\ 
Yann Alibert,\inst{\ref{inst3},\ref{inst5}}
Taras V. Gerya,\inst{\ref{inst1}}
Christoph Mordasini\inst{\ref{inst3},\ref{inst5}}
}

\institute{
Institute of Geophysics, ETH Z\"urich, Sonneggstrasse 5, Z\"urich 8092, Switzerland
\label{inst1}
\and
Bayerisches Geoinstitut, University of Bayreuth, Universit\"atsstrasse 30, Bayreuth 95440, Germany
\label{inst2} 
\and
Physikalisches Institut, University of Bern, Sidlerstrasse 5, Bern 3012, Switzerland
\label{inst3}
\and
Department of Astronomy, University of Michigan, 1085 S. University Avenue, Ann Arbor, MI 48109, United States
\label{inst4}
\and
Center for Space and Habitability, University of Bern, Gesellschaftsstrasse 6, Bern 3012, Switzerland
\label{inst5}
}

\date{February 11, 2019; preprint differing from \href{http://dx.doi.org/10.1038/s41550-018-0688-5}{journal version} (\href{https://rdcu.be/bmdlw}{free-to-read pdf})}

\abstract
{
In contrast to the water-poor inner solar system planets, stochasticity during planetary formation \citep{2017A&A...598L...5A,2017Icar..297..134R} and order of magnitude deviations in exoplanet volatile contents \citep{2017ARA&A..55..433K} suggest that rocky worlds engulfed in thick volatile ice layers \citep{2003ApJ...596L.105K,2004Icar..169..499L} are the dominant family of terrestrial analogues \citep{2015NatGe...8..177T,2018MNRAS.477.4627R} among the extrasolar planet population. However, the distribution of compositionally Earth-like planets remains insufficiently constrained, and it is not clear whether the solar system is a statistical outlier or can be explained by more general planetary formation processes. Here we employ numerical models of planet formation, evolution, and interior structure, to show that a planet's bulk water fraction and radius are anti-correlated with initial \textsuperscript{26}Al levels in the planetesimal-based accretion framework. The heat generated by this short-lived radionuclide rapidly dehydrates planetesimals \citep{1993Sci...259..653G} prior to accretion onto larger protoplanets and yields a system-wide correlation \citep{2017ApJ...849L..33M,2018AJ....155...48W} of planet bulk abundances, which, for instance, can explain the lack of a clear orbital trend in the water budgets of the TRAPPIST-1 planets \citep{2018ApJ...865...20D}. Qualitatively, our models suggest two main scenarios of planetary systems' formation: high-\textsuperscript{26}Al systems, like our solar system, form small, water-depleted planets, whereas those devoid of \textsuperscript{26}Al predominantly form ocean worlds, where the mean planet radii between both scenarios deviate by up to $\approx$10\%.
}

\maketitle

\section*{Main Text}
\label{sec:main}

In the early solar system, the decay heat from the short-lived radionuclide \textsuperscript{26}Al ($t_{\mathrm{1/2,^{26}Al}}$ $\approx$ 0.72 Myr) powered the interior evolution of planetesimals, the seeds and building blocks of the rocky planets, and led to silicate melting \citep{2014E&PSL.390..128F,2016Icar..274..350L} and degassing of primordial water abundances \citep{1993Sci...259..653G,Monteux2018}. Here, we explore the systematic effects of \textsuperscript{26}Al on rocky planetary systems using numerical models of planetary formation \citep{2014prpl.conf..691B} with \textsuperscript{26}Al-induced water loss from planetesimals during the main accretion phase \citep{Monteux2018}. We generate synthetic planet populations with internal structures defined by the planets' composition, which result in statistical variations of planet water abundance and (transit) radius.

\begin{figure*}[tbh]
    \centering
    \includegraphics[width=0.95\textwidth]{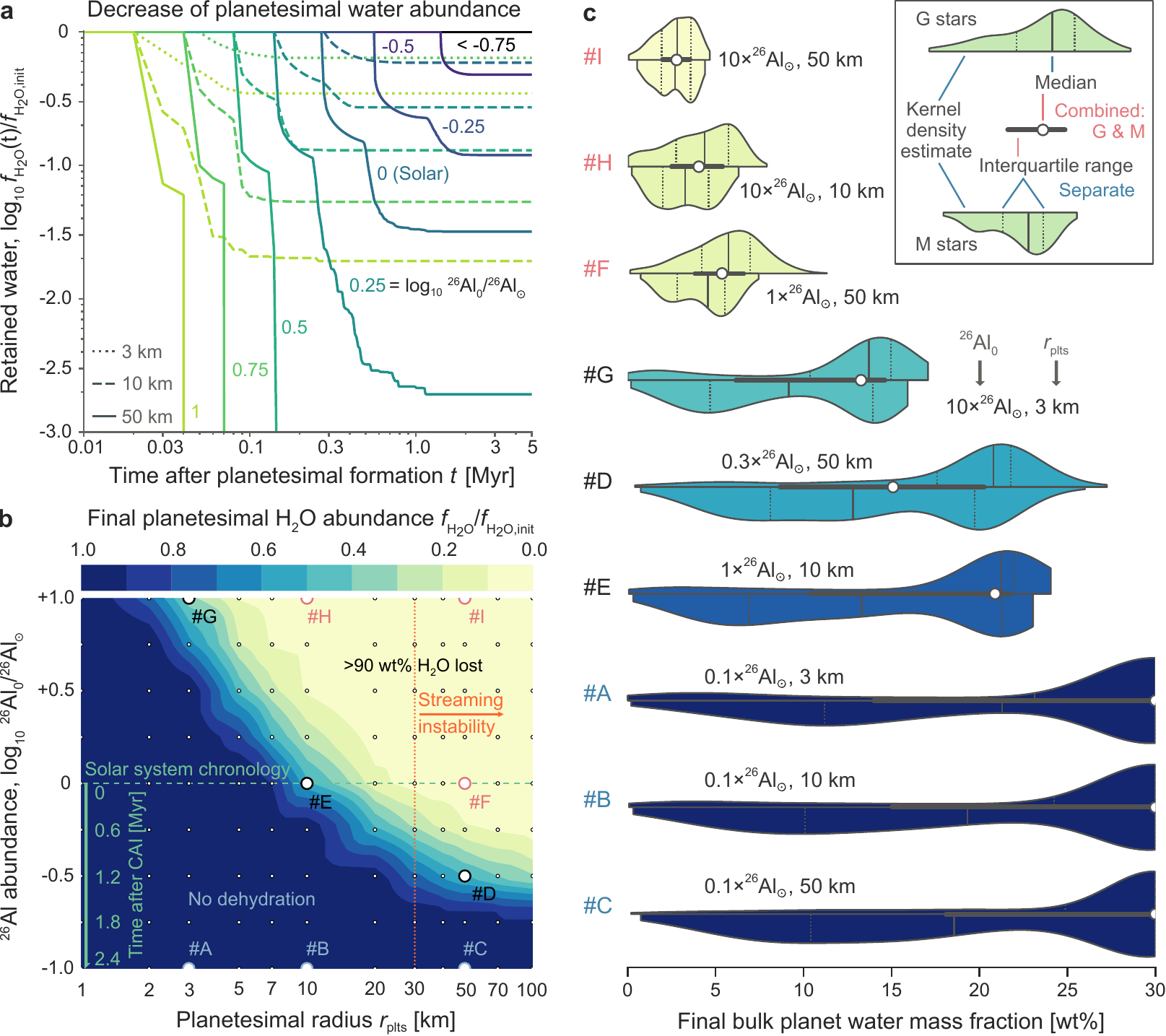}
    \caption{\textbf{Dehydration of icy planetesimals from \textsuperscript{26}Al-heating and resulting influence on planet water abundance.} \textbf{(a)} Time-resolved water retention for planetesimals of 3, 10 and 50 km radius with \textsuperscript{26}Al$_{\mathrm{0}}$ $\in$ [0.1, 10] $\times$ \textsuperscript{26}Al$_{\odot}$. Brighter colors indicate stronger water depletion. The degassing saturation, when the lines become horizontal, results from the rapid decay of \textsuperscript{26}Al. \textbf{(b)} Final state of water retention. \textsuperscript{26}Al$_{\mathrm{0}}$ at planetesimal formation can be translated into planetesimal formation time after CAIs for solar system objects. The orange line depicts the approximate lowest mass planetesimals inferred for the early solar system planetesimal population \citep{2017Sci...357.1026D}. \textbf{(c)} Distribution and shift in planet bulk water abundances for specific planetesimal configurations from \textbf{a} and \textbf{b}, for planet masses $M_{\mathrm{P}}$ $\in$ [0.1, 10] $\times$ $M_{\mathrm{Earth}}$ and \textit{f}$_{\mathrm{H_{2}O}}$ $>$ 0. The legend for the violin distributions is given in the upper-right box. Each configuration shows the statistical distribution of \textit{f}$_{\mathrm{H_{2}O}}$ in a synthetic planet population generated from our model. The white dot in the gray bar in the middle of each violin histogram represents the median of the entire (combined G and M star) planet population, the horizontal gray bar the interquartile range (middle 50\% of the population within the bar, upper and lower 25\% outside the bar), again combined. The vertical solid and dashed lines in the upper and lower violin histogram represent the median and interquartile range, respectively, for the G or M star planet population in isolation. The water retention in planetesimals from \textbf{a} and \textbf{b} is correlated with the final retained water in \textbf{c} (color scales in \textbf{b} and \textbf{c} are equal).}
    \label{fig:fig1}
\end{figure*}

In the models presented, initially Moon-sized protoplanets grow from the accretion of 1--100 km-sized planetesimals and gas, and migrate within the protoplanetary disk of G or M-type systems (\nameref{sec:methods}). The initial location of the embryos and the starting disk structures and boundaries are randomized to reflect the diversity found in observed young planetary systems \citep{2016ApJ...828...46A}. Planetesimals are set to be dry within the snowline and icy outside, with a decreasing water mass fraction over time, calculated from planetesimal interior models that account for the dehydration from internal radiogenic heating of \textsuperscript{26}Al. Here, we account for accretion of planetesimals only, and ignore the potential contribution from smaller particles, such as pebbles \citep{2015SciA....115109J,2018Natur.555..507S,Alibert18NatAstron}. The heating rate in the planetesimal interior is controlled by the amount of \textsuperscript{26}Al incorporated upon planetesimal formation, which may vary substantially between planetary systems \citep{2016MNRAS.462.3979L,2018PrPNP.102....1L}. We account for this variability by generating synthetic planet populations with different planetesimal radii, $r_{\mathrm{plts}}$ = 3, 10, 50 km, and initial \textsuperscript{26}Al abundances of \textsuperscript{26}Al$_{\mathrm{0}} \in$ [0.1, 10] $\times$ \textsuperscript{26}Al$_{\odot}$, with \textsuperscript{26}Al$_{\odot}$ the solar system's `canonical' (\textsuperscript{26}Al/\textsuperscript{27}Al)$_{\mathrm{0}}$ at CAI formation, and compare them to a nominal case without \textsuperscript{26}Al-heating (for further details on the models see \nameref{sec:methods}). For each combination of $r_{\mathrm{plts}}$, \textsuperscript{26}Al$_{\mathrm{0}}$ and stellar type (G or M) we performed 30'000 single planet simulations \citep{2017A&A...598L...5A,2014prpl.conf..691B}, resulting in a statistically representative set of 540'000 individual simulations over 18 parameter sets (cf. Fig. \ref{fig:fig1}).

The control of \textsuperscript{26}Al$_{\mathrm{0}}$ and $r_{\mathrm{plts}}$ on the retention of water within planetesimals and resulting planet populations from a given set of initial conditions are shown in Fig. \ref{fig:fig1}. Planetesimals with larger $r_{\mathrm{plts}}$ and higher \textsuperscript{26}Al$_{\mathrm{0}}$ dehydrate faster and up to 100\% for extreme values. Rooted in our conservative choice for dehydration (\nameref{sec:methods}), the total water loss divides the parameter range into two distinct regimes. The first consists of almost pristine water-rock ratios for small planetesimals with low \textsuperscript{26}Al$_{\mathrm{0}}$. However, for \textsuperscript{26}Al$_{\mathrm{0}}$ $\gtrsim$ \textsuperscript{26}Al$_{\odot}$ and planetesimals with $r_{\mathrm{plts}}$ $\gtrsim$ 10 km, water loss is nearly complete (Fig. \ref{fig:fig1}a,b).

\begin{figure*}[tbh]
    \centering
    \includegraphics[width=1.0\textwidth]{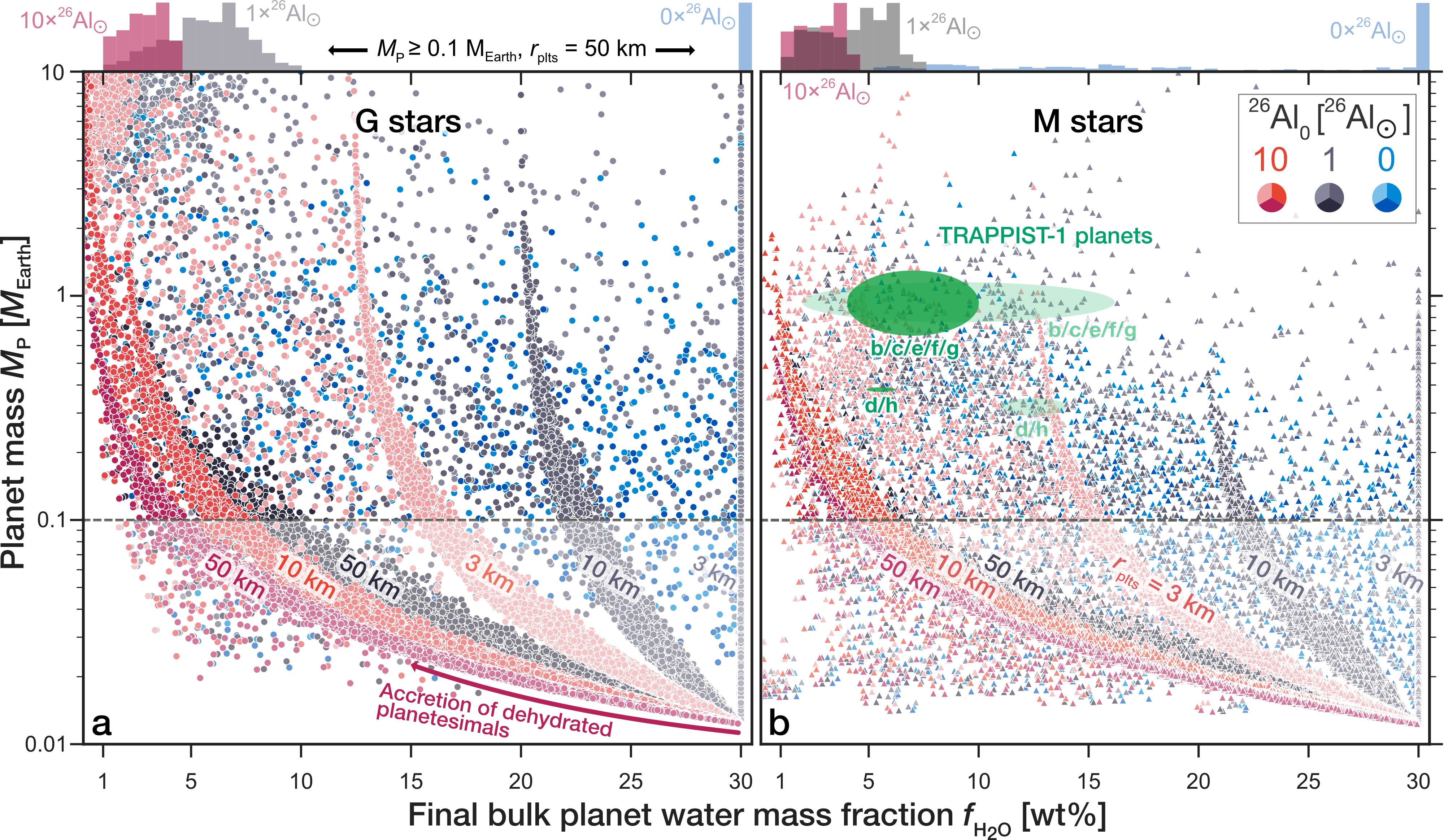}
    \caption{
    \textbf{Gradual desiccation of protoplanets as a function of \textsuperscript{26}Al\textsubscript{0} for planets with \textit{f}\textsubscript{H\textsubscript{2}O} $>$ 0.} For increasing \textsuperscript{26}Al$_{\mathrm{0}}$ and $r_{\mathrm{plts}}$, the bulk planet water fraction $f_{\mathrm{H_{2}O}}$ decreases systematically. For $M_{\mathrm{P}}$ $\geq$ 0.1 $M_{\mathrm{Earth}}$ and $r_{\mathrm{plts}}$ = 50 km, the $f_{\mathrm{H_{2}O}}$ histogram on top shows approximately one order of magnitude deviation between planets formed devoid of \textsuperscript{26}Al and with \textsuperscript{26}Al $\gtrsim$ \textsuperscript{26}Al$_{\odot}$. Only the latter cases increasingly populate the terrestrial planet regime with $f_{\mathrm{H_{2}O}}$ $\lesssim$ \ensuremath{\mathcal{O}}{(wt\%)}. The blue points from the populations with zero \textsuperscript{26}Al$_{\mathrm{0}}$ only rarely and stochastically form planets with low water mass fractions due to rapid inward migration. The clustering for \textsuperscript{26}Al$_{\mathrm{0}}$ = 0 at the maximum water mass fraction is inherited from the chosen initial composition of planetesimals beyond the snowline (\nameref{sec:methods}). It is important to note that the areas of clustering locate the \emph{maximum} water mass fractions for a given planet mass within a synthetic population, i.e., planets that are formed entirely beyond the snowline. For example, all planets from the synthetic population with $r_{\mathrm{plts}}$ = 3 km, and \textsuperscript{26}Al$_{\mathrm{0}}$ = 10 $\times$ \textsuperscript{26}Al$_{\odot}$, show water mass fractions $f_{\mathrm{H_{2}O}} \lesssim$ 15 wt\% for $M_{\mathrm{P}}$ $\geq$ 0.1 $M_{\mathrm{Earth}}$. G stars (\textbf{a}) on average form higher-mass planets than M stars (\textbf{b}) because of their higher initial total mass budget in the disk. The TRAPPIST-1 planets are shown as inferred by \cite{2018ApJ...865...20D} for the planet masses from \cite{2018A&A...613A..68G} (light green) and with potential systematic shifts in the data accounted for (dark green). They are consistent with being formed in a planetary system with \textsuperscript{26}Al$_{\mathrm{0}}$ $\gtrsim$ \textsuperscript{26}Al$_{\odot}$ and $r_{\mathrm{plts}}$ $\gtrsim$ 10 km and plot in sparsely populated regions of the \textsuperscript{26}Al = 0 planet populations.
    }
    \label{fig:fig2}
\end{figure*}

For distinct combinations of \textsuperscript{26}Al$_{\mathrm{0}}$ and $r_{\mathrm{plts}}$, we simulate the influence on the expected planet population for planet masses $M_{\mathrm{P}}$ $\in$ [0.1, 10] $\times$ $M_{\mathrm{Earth}}$ (Fig. \ref{fig:fig1}c). Because the timescale for water-loss caused by \textsuperscript{26}Al-heating is significantly shorter than the accretion timescale, sufficiently \textsuperscript{26}Al-enriched planetesimals are mostly dry when they accrete onto protoplanets. Therefore, the final planet water mass fractions are correlated with the retained water fraction in planetesimals due to \textsuperscript{26}Al-heating. The planet desiccation caused by the accretion of ever-more dehydrated planetesimals reduces the inherent scatter and range in $f_{\mathrm{H_{2}O}}$ in the synthetic planet populations (cf. Fig. \ref{fig:fig1}c and \ref{fig:fig2}a,b). For fixed planetesimal radius and increasing \textsuperscript{26}Al$_{\mathrm{0}}$, accreting planets receive more relative mass contribution from dry objects and end up water-depleted relative to nominal conditions. 

In the solar system, the initial planetesimal size frequency distribution is expected to have been dominated by bodies with $r_{\mathrm{plts}}$ $\gtrsim$ 30--50 km  \citep{2017Sci...357.1026D}. For such bodies, the equilibrium between radiogenic heating and surface cooling stabilizes internal temperatures for an extended timespan at spatially isothermal conditions \citep{2016Icar..274..350L}. Therefore, the fractional dehydration in Fig. \ref{fig:fig1}b flattens above $r_{\mathrm{plts}}$ $\gtrsim$ 50 km and becomes nearly independent of planetesimal size. For $r_{\mathrm{plts}}$ = 50 km, dehydration is dominantly controlled by \textsuperscript{26}Al$_{\mathrm{0}}$ and generates a dichotomy between planets in \textsuperscript{26}Al-enriched (top black and red histograms, Fig. \ref{fig:fig2}) versus non-enriched systems (top blue histogram, Fig. \ref{fig:fig2}). M and G stars overall display a similar trend, but M stars form smaller planets on average, due to their lower initial budget of planet-forming material.

\begin{figure}[tb!]
    \resizebox{\hsize}{!}{\includegraphics{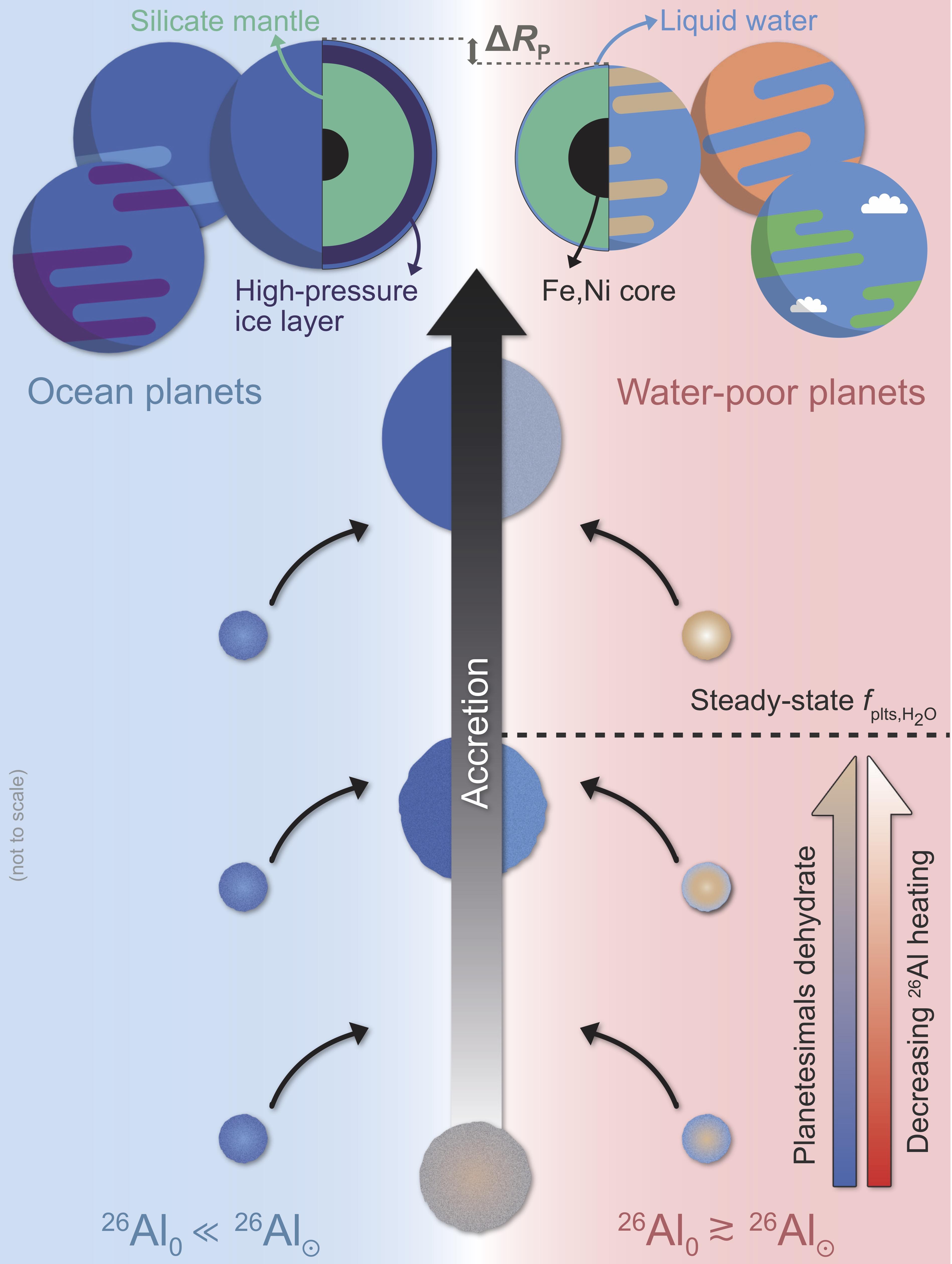}}
    \caption{\textbf{Qualitative sketch of the effects of \textsuperscript{26}Al enrichment on planetary accretion and the deviations between \textsuperscript{26}Al-poor (left) and \textsuperscript{26}Al-rich (right) planetary systems.} Arrows indicate proceeding accretion (middle), planetesimal water content (bottom right, blue-brown), and live \textsuperscript{26}Al (bottom right, red-white).}
    \label{fig:fig3}
\end{figure}

The emerging trend from our simulations is illustrated in Fig. \ref{fig:fig3}, with a clear distinction between planetary systems that are significantly enriched (\textsuperscript{26}Al$_{\mathrm{0}}$ $\gtrsim$ \textsuperscript{26}Al$_{\odot}$), and those that are not. In general, \textsuperscript{26}Al is expected to be abundant but inhomogeneously distributed within young star-forming regions \citep{2016MNRAS.462.3979L,2016ApJ...826...22K}. According to our simulations, planets in enriched systems grow from ever-more dehydrated planetesimals and form desiccated planets in their terrestrial planet zone. Depending on the initial planetesimal sizes, final planet water fractions are up to two orders of magnitude below the initial planetesimal water mass fractions, and are strongly correlated with the efficency of dehydration during accretion (Fig. \ref{fig:fig1}).

\begin{figure}[tb!]
    \resizebox{\hsize}{!}{\includegraphics{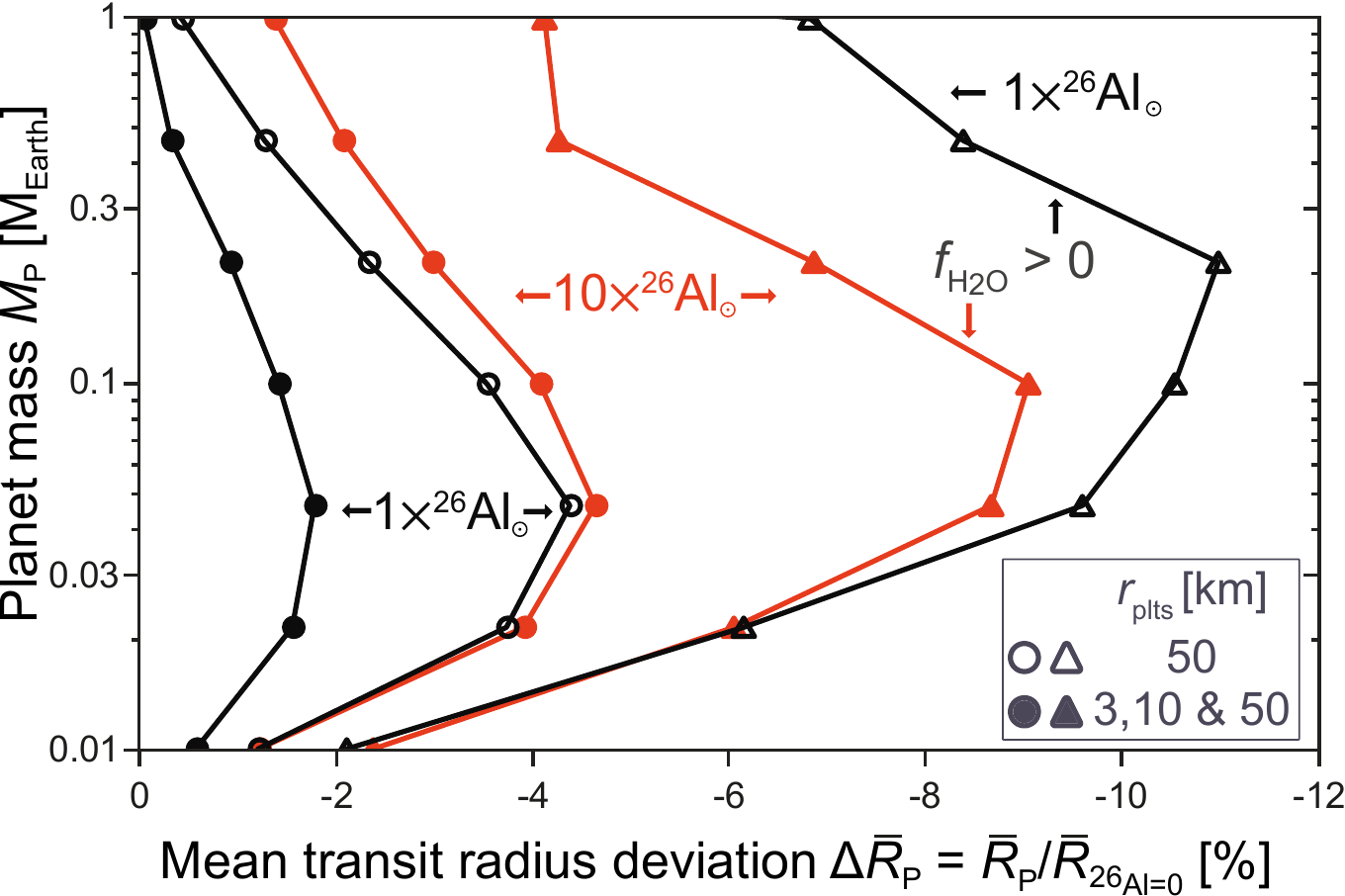}}
    \caption{\textbf{Shift in mean transit radii between planetary populations with varying \textsuperscript{26}Al enrichment.} Quantitative predictions for planetary mean transit radii (\nameref{sec:methods}) for given mass bins in planetary systems with \textsuperscript{26}Al, normalized to the case completely devoid of \textsuperscript{26}Al. Open symbols indicate radius deviations averaged over simulations with $r_{\mathrm{plts}}$ = 50 km, filled symbols combine 3, 10 and 50 km planetesimals. Triangles select for planets with \textit{f}$_{\mathrm{H_{2}O}}$ $>$ 0, circles include also completely dry planets that formed entirely inside of the water snowline. For varying selection criteria, the planet radii per mass bin for \textsuperscript{26}Al-enriched systems deviate from non-enriched systems by up to $\approx$10 \%.}
    \label{fig:fig4}
\end{figure}

The bulk volatile mass fraction has the greatest influence on the structure and mass-radius relation of a rocky planet \citep{2017SSRv..212..877N,2018NatAs...2..297U}. Therefore, we anticipate the resulting smaller radii (from lower water mass fraction) for higher \textsuperscript{26}Al levels to be reflected in the galactic exoplanet population. For deviations in planet bulk water fractions predicted here, the thickness of the volatile layer on top of the silicate mantle constitutes several per cent of the radius \citep{2003ApJ...596L.105K,2004Icar..169..499L,2014A&A...561A..41A,2017SSRv..212..877N}. We calculate this deviation in our synthetic populations by translating the derived planetary masses and compositions into a mean radius in a given mass bin (Fig. \ref{fig:fig4}) using interior structure models that are sensitive to the total planet mass, its water and (captured) hydrogen/helium mass fraction, and the surface pressure (\nameref{sec:methods}). For the entire populations of planets among G and M stars, the radius deviation reaches up to 2\% for 1 $\times$ \textsuperscript{26}Al$_{\odot}$, and can go up to $\approx$4\% deviation for Mars-sized planets for 10 $\times$ \textsuperscript{26}Al$_{\odot}$ or $r_{\mathrm{plts}}$ = 50 km, respectively. If we only consider planets that accrete a minimum amount of water (planets that receive some mass contribution from beyond the iceline), with 10 $\times$ \textsuperscript{26}Al$_{\odot}$, or 1 $\times$ \textsuperscript{26}Al$_{\odot}$ with $r_{\mathrm{plts}}$ = 50 km, the mean-radius shift can reach up to $\approx$10\%. Planetary systems with high \textsuperscript{26}Al$_{\mathrm{0}}$ (\textsuperscript{26}Al$_{\mathrm{0}}$ $\gtrsim$ 1--10 $\times$ \textsuperscript{26}Al$_{\odot}$) form water-depleted planets and with system-wide smaller radii compared to the non-enriched population.

Such deviations are expected to be measurable by the planned PLATO mission \citep{2014ExA....38..249R}, which will aim to characterize a statistical ensemble of planetary radii in the rocky planet regime. The intrinsic compositional scatter in the inferred mean densities from known exoplanets suggests a large stochastic component in the planet formation process. Yet, recent analyses of data based on Kepler multi-planet systems provided strong evidence for intra-system correlation between planetary radii \citep{2017ApJ...849L..33M,2018AJ....155...48W}. Therefore, in the exoplanet census probed so far, the fate and long-term structure of planets seems to be dominated by physical and chemical effects on a system-to-system level, rather than emerging from intra-system stochasticity during accretion, such as impact stripping \citep{2010ApJ...719L..45M,2016ApJ...817L..13I}. With future access to a statistical ensemble of low-mass planet radii from exoplanet-focused missions, the highly \textsuperscript{26}Al-enriched systems, such as the solar system, where planetary radii deviate by several per cent from the population of \textsuperscript{26}Al-poor systems, may stick out from the mean of the population and provide clues about the underlying \textsuperscript{26}Al distribution of planetary systems.

For example, the system-wide water depletion of the TRAPPIST-1 planets \citep{2018A&A...613A..68G,2018ApJ...865...20D} is consistent with \textsuperscript{26}Al $\gtrsim$ \textsuperscript{26}Al$_{\odot}$-induced desiccation (Fig. \ref{fig:fig2}). The atmospheres of the TRAPPIST-1 planets seem to be secondary \citep{2018NatAs...2..214D}, and may have lost several Earth ocean equivalents of water \citep{2017AJ....154..121B}. However, to account for the consistency of especially the outermost planets \emph{e-h} with near Earth-like volatile abundances and lack of an orbital trend in water budget \citep{2018ApJ...865...20D}, an order of magnitude depletion mechanism, such as suggested here, must affect all of the planets. Therefore, the retrieved low water mass fractions of the TRAPPIST-1 \citep{2018A&A...613A..68G,2018ApJ...865...20D} planets are unexpected from formation and evolution models \citep{2015NatGe...8..177T,2015ApJ...804....9C,2017A&A...604A...1O,2017A&A...598L...5A,2018NatAs...2..297U}, and present a severe challenge for current planet formation scenarios. The \textsuperscript{26}Al desiccation mechanism we put forward achieves system-wide water depletion for G and M stars without the need to fit specific accretion dynamics, as it has been proposed \citep{2017A&A...604A...1O,2018NatAs...2..297U}. However, because the TRAPPIST-1 system is just the first of perhaps many such systems, coordinated observational efforts will be required to establish population characteristics for similar systems, in order to distinguish between \textsuperscript{26}Al desiccation and migration-driven mechanisms \citep{2017A&A...604A...1O,2018NatAs...2..297U} as the origin of the TRAPPIST-1 planet compositions.
 
For accretion scenarios where planetesimals represent the primary carrier of water, our models suggest that planetary systems with \textsuperscript{26}Al abundances similar to or higher than the solar system generically form terrestrial planets with low water mass fractions, $f_{\mathrm{H_{2}O}}$ $\lesssim$ \ensuremath{\mathcal{O}}{(wt\%)}. This effect is more pronounced for planets further out from their host star, as embryos in these regions grow preferentially from water-rich solids. For a non-uniform distribution of \textsuperscript{26}Al in Milky Way star-forming regions, the systematic water depletion in \textsuperscript{26}Al-enriched systems suggests the existence of two qualitatively distinct classes of planetary systems: water-poor (\textsuperscript{26}Al-rich) and water-rich (\textsuperscript{26}Al-poor) systems, with a systematic mean-radius deviation for sub-Earth terrestrial planets between these classes. The resulting shape of the distribution of dry and wet planetary systems depends on the genuine, but unknown, distribution of \textsuperscript{26}Al levels \citep{2018PrPNP.102....1L} among planet-forming systems and the nature and timing of protoplanet accretion. If rocky planets grow primarily from the accumulation of planetesimals, then the suggested deviation between planetary systems should be clearly distinguishable among the rocky exoplanet census. If, however, the main growth of rocky planets may proceed from the accumulation of small particles, such as pebbles, then the deviation between \textsuperscript{26}Al-rich and \textsuperscript{26}Al-poor systems may become less clear, and the composition of the accreting pebbles needs to be taken into account. Therefore, in future work, models of water delivery and planet growth need to synchronize the timing of earliest planetesimal formation \citep{2018A&A...614A..62D}, the mutual influence of collisions \citep{2018Icar..302...27L} and \textsuperscript{26}Al dehydration, the potential growth by pebble accretion \citep{2015SciA....115109J,2018Natur.555..507S,Alibert18NatAstron}, and the partitioning of volatile species between the interior and atmosphere of growing protoplanets \citep{2018SSRv..214...76I} in order to further constrain the perspectives for rocky exoplanet evolution \citep{2018ApJ...864...75K}.

\section*{Methods}
\label{sec:methods}

\subsection*{Planetesimal dehydration}
\label{sec:internal}

We model water loss from instantaneously-formed planetesimals composed of a rock-ice mixture using numerical models that employ a conservative finite-differences, fully-staggered grid method coupled to a marker-in-cell approach \citep{2007PEPI..163...83G,2014M&PS...49.1083G}. The thermo-chemical evolution of planetesimals is computed in a two-dimensional infinite cylinder geometry on a Cartesian grid, solving the Poisson, continuity, Stokes and energy conservation equations. We assume the planetesimals to be accreted with the temperature of the protoplanetary disk beyond the water snowline, $T_0 = 150$ K, which is kept constant during the evolution of the planetesimal utilizing the free-surface `sticky-air' method \citep{2012GeoJI.189...38C}. Heating is provided by the decay of \textsuperscript{26}Al, which defines the radiogenic heat source term over time 
\begin{align}
	H_{^{26}\mathrm{Al}}(t) = f_{\mathrm{Al}} \cdot (^{26}\mathrm{Al}/^{27}\mathrm{Al})_0 \cdot E_{\mathrm{^{26}\mathrm{Al}}} \cdot \mathrm{exp}({-t/\tau_{^{26}\mathrm{Al}}}) / \tau_{^{26}\mathrm{Al}},
\end{align}
with the chondritic abundance of aluminum, $f_{\mathrm{Al}}$ \citep{2003ApJ...591.1220L}, the ratio of \textsuperscript{26}Al to stable $^{27}$Al at the time of planetesimal formation, \textsuperscript{26}Al$_{\mathrm{0}}$ = (\textsuperscript{26}Al/$^{27}$Al)$_0$, the decay energy, $E_{\mathrm{^{26}\mathrm{Al}}} = 3.12$ MeV \citep{2009Icar..204..658C}, and the mean lifetime, $\tau_{^{26}\mathrm{Al}} = 1.03$ Myr. We ignore any potential heat contribution from $^{60}$Fe, which may further boost radiogenic heating rates in extrasolar systems \citep{2016MNRAS.462.3979L,2017MNRAS.464.4318N}. If the planetesimal interior reaches temperatures beyond the rock disaggregation threshold \citep{2009GGG....10.3010C} at a silicate melt fraction of $\phi \gtrsim 0.4$, where the rock viscosity drops by more than ten orders of magnitude, we approximate the thermal conductivity in the soft turbulence limit \citep{1994AnRFM..26..137S} with
\begin{align}
k_{\mathrm{eff}} = (q/0.89)^{3/2} \cdot \alpha_{\mathrm{liq}} g c_{\mathrm{p}} / (\Delta T^2 \rho_{\mathrm{s}} \eta_{\mathrm{num}}),
\end{align}
with the convective heat flux, $q$, the temperature difference across nodes, $\Delta T$, silicate density, $\rho_{\mathrm{s}}$, thermal expansivity of molten silicates, $\alpha_{\mathrm{liq}}$, silicate heat capacity, $c_{\mathrm{p}}$, local gravity, $g(x,y)$, and lower cut-off viscosity, $\eta_{\mathrm{num}}$. For numerical values used and further details and references on the code see \cite{2016Icar..274..350L}. The initial planetesimal water-to-rock ratio beyond the snowline is expected to be between $\approx$0.05 \citep{2015ApJ...804....9C,2018SSRv..214...47O}, the water content of carbonaceous chondrites, and $\approx$0.5, as suggested by equilibrium condensation calculations \citep{2003ApJ...591.1220L}. Here, we adopt a value closer to the upper estimate, $f_{\mathrm{H_{2}O,init}}$ = 0.3, but our calculations only marginally depend on the adopted value. 

In general, during heat-up of a primordial water ice-rock mixture, ices melt and react with the ambient rock. The liquid water undergoes pore water convection and escapes quickly once the gas phase is reached \citep{1993Sci...259..653G}, but a small fraction of water may be trapped in hydrous silicate phases. Therefore, we numerically account for dehydration of parts of the planetesimal interior at a conservative upper limit of $T \geq T_{\mathrm{dry}}$ = 1223 K, the upper limit of the amphibolite stability field, when any possibly remaining hydrous silicate phases break down. At these high temperatures exsolved water vapor is lost quasi-instantaneously because planetesimals of this size cannot preserve an outgassed atmosphere. We do not resolve potential earlier water-loss from degassing \citep{castillorogez2017}, residual volatiles above $T_{\mathrm{dry}}$ \citep{fu2017}, or ice sublimation during late and optically thin disk stages \citep{2010ApJ...716.1252M}. Using these assumptions, we compute the expected ratio of dehydrated to primordial water-rock mixture at time $t$ due to degassing, 
\begin{align}
f_{\mathrm{H_{2}O}}(t)/f_{\mathrm{H_{2}O,init}} = 1 - X_{\mathrm{dry}}(t)/X_{\mathrm{plts}},
\end{align}
with the dry fraction, $X_{\mathrm{dry}}(t)$, of the total planetesimal interior, $X_{\mathrm{plts}}$, and the initial water-to-rock ratio, $f_\mathrm{H_{2}O,init}=0.3$. Under these conditions, a planetary system in the planet formation model is represented by an initial \textsuperscript{26}Al$_{\mathrm{0}}$ that corresponds to the time of planetesimal formation. If the \textsuperscript{26}Al content may vary spatially within the disk, as it was suggested \citep{2011ApJ...735L..37L,2015E&PSL.420...45S}, the solar system itself would be represented by a sub-canonical (\textsuperscript{26}Al$_{\mathrm{0}}$ $\leq$ \textsuperscript{26}Al$_{\odot}$) value, similar to the effects of delayed planetesimal formation (Fig. \ref{fig:fig1}b, cf. \cite{2018Icar..302...27L} for a discussion of the effects on planetesimal evolution).

\subsection*{Planet formation}

We compute the formation of planets and generate our synthetic planet populations using an updated version of the model of \cite{2005A&A...434..343A}. The computer code numerically treats the structure and evolution of the protoplanetary disk, the dynamical properties and accretion rate of planetesimals onto accreting protoplanets, the planetary envelope structure, and disk-planet interactions \citep{2009A&A...501.1139M,2009A&A...501.1161M,2015IJAsB..14..201M,2014prpl.conf..691B}. Here, we provide a brief summary of the most important code modules used in this work.

The protoplanetary disk model relies on the Shakura-Sunyaev \citep{1973A&A....24..337S} disk viscosity approximation ($\alpha_{\mathrm{disk}} = 2 \times 10^{-3}$) and computes the surface density evolution over time by solving the radial diffusion equation,
\begin{align}
\frac{\mathrm{d} \Sigma}{\mathrm{d} t} = \frac{3}{r} \frac{\partial}{\partial r} \left[ r^{1/2} \frac{\partial}{\partial r} \tilde{\nu} \Sigma r^{1/2} \right]+ \dot{\Sigma}_{\mathrm{w}} + \dot{Q}_{\mathrm{planet}},
\end{align}
with the surface density, $\Sigma$, orbital radius, $r$, effective viscosity, $\tilde{\nu}$, and gas accretion onto embryos, $\dot{Q}_{\mathrm{planet}}$, calculated from removing gas in an annulus centered on the embryo with a width of one Hill radius, 
\begin{align}
R_{\mathrm{H}} = a_{\mathrm{planet}} \left[ M_{\mathrm{planet}} / (3 M_{\mathrm{star}}) \right]^{1/3},
\end{align}
with the planet semi-major axis, $a_{\mathrm{planet}}$, planet mass, $M_{\mathrm{planet}}$, and star mass, $M_{\mathrm{star}}$. Mass loss due to internal (EUV) photoevaporation $\dot{\Sigma}_{\mathrm{w}}$ \citep{Clarke2001} is set $\propto r^{-5/2}$ outside of a gravitational radius of $\approx 5$ au and external (FUV) photoevaporation \citep{2003ApJ...582..893M} is constant outside of $\approx 140$ au, with the total mass loss being a free model parameter. The model used to represent the planetesimal disk relies on the initial central temperature and pressure from the disk model to compute the location of the water snowline, thereby neglecting radial drift of planetesimals \citep{2000ApJ...528..995S} and that of the snowline. Drift timescales for planetesimals larger than 1 km exceed the disk lifetime by orders of magnitude \citep{1977MNRAS.180...57W}. 

We consider rocky planetesimals ($\overline{\rho}_{\mathrm{plts-dry}} = \rho_{\mathrm{rock}}$ = 3200 kg/m$^3$) inside, and rock-ice aggregates ($\overline{\rho}_{\mathrm{plts-ice,init}} = \rho_{\mathrm{H_2O}} f_{\mathrm{H_2O,init}} + \rho_{\mathrm{rock}}[1-f_{\mathrm{H_2O,init}}]$) beyond the snowline, which are fixed in radius and accrete onto the planetary embryo that is embedded in the disk in a single simulation. The residual water mass fraction, $f_\mathrm{H_{2}O}(t)$, of the accreting planetesimals is computed from the internal evolution (\nameref{sec:internal}) and is translated to a decreasing planetesimal density, $\overline{\rho}_\mathrm{plts-ice}(t)$, and disk solid surface density, $\Sigma_\mathrm{plts-ice}(t)$, by reducing the planetesimal density as 
\begin{align}
\overline{\rho}_\mathrm{plts-ice}(t) =\rho_\mathrm{H_{2}O}  f_\mathrm{H_{2}O}(t) +\rho_\mathrm{rock}[1-f_\mathrm{H_{2}O,init}].
\end{align}
In this formulation, lost water is assumed to be replaced by pore space and the planetesimal radius stays constant. The solid surface density available for embryos to accrete beyond the iceline thus changes with
\begin{align}
\Sigma_\mathrm{plts-ice}(t) = \Sigma_{\mathrm{plts-ice,init}} \cdot \left( f_\mathrm{H_{2}O}(t) + [1-f_\mathrm{H_{2}O,init}]\right).
\end{align}

In our nominal model, a single embryo of initially lunar mass, $M=0.0123$ $M_{\mathrm{Earth}}$, is placed randomly between specific inner and outer bounds within the protoplanetary disk (\nameref{sec:parameter_space}), with a dry composition inside the snowline, and wet outside. It starts accreting solids (planetesimals) and gas, and may migrate in the type I and II regime, depending on the embryo mass and physical structure of the disk at a given orbit \citep{2014A&A...567A.121D}. The solid accretion rate \citep{2001Icar..149..235I,2003A&A...410..711I} takes into account the captured atmosphere. Planetesimal excitation and damping is computed by taking into account self interactions and damping by gas drag \citep{2013A&A...549A..44F}. We ignore water loss due to collisions, which may further reduce the water inventory \citep{2003Icar..164..149G,2015Icar..247...81S,2016ApJ...817L..13I,2018CeMDA.130....2B} dependent on the frequency of such interactions, and accretion of smaller solid particles \citep{2015SciA....115109J,2017SciA....3E0407B,2018Natur.555..507S} ('pebbles') that may shift the ratio of dry to wet accreted primitive materials \citep{2016E&PSL.452...36C}. Gas accretion due to planetary contraction is considered using a dust opacity reduction factor of $0.01$ compared to interstellar values \citep{1996Icar..124...62P,2013A&A...558A.109A}.

G star settings are identical to the Sun's values. The properties for the M star runs are scaled down. We choose a fixed mass of $M_\mathrm{star} = 0.2 \; M_{\odot}$ for the M stars. The radius of the star is set to
\begin{align}
R_\mathrm{star} = (M_\mathrm{star}/M_{\odot})^{0.945} R_{\odot},
\end{align}
with luminosity \citep{1991Ap&SS.181..313D}
\begin{align}
 L_\mathrm{star} = 0.628 (M_\mathrm{star}/M_{\odot})^{2.62} L_{\odot},
\end{align} 
and temperature
\begin{align}
T_\mathrm{star} = \sqrt[4]{L_\mathrm{star}/(4\pi R_\mathrm{star}^2 \sigma)},
\end{align}
with stellar radius, $R_\mathrm{star}$, stellar mass, $M_\mathrm{star}$, and Stefan-Boltzmann constant, $\sigma$. The disk dimensions, exponential cut-off radius and the embryo placement boundaries (\nameref{sec:parameter_space}) are reduced to account for the lower masses and sizes of M star disks. Thus, initially all embryos form closer to the star compared to the G star populations. The initial disk mass follows the scaling law \citep{2011A&A...526A..63A}
\begin{align}
M_\mathrm{disk} \propto (M_\mathrm{star}/M_{\odot})^{1.2},
\end{align}
with the internal photoevaporation rate adapted to match similar mean lifetimes compared to the G star simulations. In reality, these could be anti-correlated with stellar mass \citep{2009ApJ...695.1210K}, which would increase the efficiency of the \textsuperscript{26}Al-dehydration mechanism for M stars due to longer accretion timescales.

\subsection*{Interior structure \& evolution}
The interior structure and the long-term evolution of the planets is calculated as described in \cite{2012A&A...547A.112M} by solving the classical one-dimensional radially symmetric interior structure equations of mass conservation, hydrostatic equilibrium, and energy transport \citep{1986Icar...67..391B}
\begin{alignat}{2}\label{eq:internalstruct}
\frac{\partial m}{\partial r_{\mathrm{p}}}&=4 \pi r_{\mathrm{p}}^{2} \rho,\\
\quad  \quad \frac{\partial P}{\partial r_{\mathrm{p}}}&=-\frac{G m}{r_{\mathrm{p}}^{2}}\rho,    \\
\frac{ \partial T}{\partial r_{\mathrm{p}}}&=\frac{T}{P}\frac{\partial P}{\partial r_{\mathrm{p}}}\nabla(T,P),
\end{alignat}
where $r_{\mathrm{p}}$ is the radial distance from the planet's center, $m$ the enclosed mass, $P$ the pressure, $\rho$ the density,  and $G$  the gravitational constant. The intrinsic luminosity of a planet is assumed to be constant as a function of the planet radius. The gradient $\nabla$ depends on the process by which the energy is transported (radiative diffusion or convection).
These calculations yield the radii of the planets given their mass and bulk composition, namely the mass fractions of iron, silicates, water, and H/He, as an output from the planet formation and planetesimal dehydration models. For the H/He envelope, the equation of state of \cite{saumonchabrier1995} is used to solve the structure equations, while for the solid part of the planet, including the water content, the modified polytropic equations of state of \cite{2007ApJ...669.1279S} are employed. The transit radius is estimated as in \cite{guillot2010}.

The loss of the primordial H/He envelope by atmospheric escape is considered in the energy- and radiation-recombination-limited approximation as described in \cite{2014ApJ...795...65J} and results in the loss of the primary atmosphere for low-mass planets at smaller orbital distances. Because of the limited water solubility in silicate mantles, the radius of planets without primordial H/He envelopes depends strongly on the water mass fraction \citep{jinmordasini2018}, and thus reveals the dehydration pattern caused by different contents of \textsuperscript{26}Al. Here, we do not treat interior-atmosphere exchange during early magma ocean phases that may further fractionate the volatile distribution within the body, in particular for close-in planets \citep{2018SSRv..214...76I}.

\subsection*{Parameter space}
\label{sec:parameter_space}

From $\gamma$-ray observations there is evidence for a widespread and heterogeneous distribution of \textsuperscript{26}Al in the galaxy \citep{2018PrPNP.102....1L}. Observational evidence from young star-forming regions and theoretical work suggest a non-uniform enrichment pattern among planetary systems \citep{2014ApJ...789...86A,2015A&A...582A..26G,2015PhyS...90f8001P,2016MNRAS.462.3979L,2016ApJ...826...22K,2017MNRAS.472L..75P,2017MNRAS.464.4318N,2017ApJ...851..147D,2018PrPNP.102....1L} with order of magnitude deviations from the solar system's `canonical' \textsuperscript{26}Al value \citep{2013M&PS...48.1383K} of \textsuperscript{26}Al$_{\odot}$ = (\textsuperscript{26}Al/$^{27}$Al)$_{\mathrm{0}}$ $= 5.25 \times 10^{-5}$. To account for these variations, we consider values in the range \textsuperscript{26}Al$_{\mathrm{0}}$ $\in$ [0.1, 10] $\times$ \textsuperscript{26}Al$_{\odot}$. In addition to initial \textsuperscript{26}Al abundance, the radii of planetesimals during accretion yield different thermal evolutionary sequences and thus dehydration patterns \citep{2016Icar..274..350L,Monteux2018}. Here, we test values in the range $r_{\mathrm{plts}}$ $\in$ [1, 100] km. However, we note that from asteroid-belt inferences and numerical studies of the streaming instability mechanism, radii larger than $\gtrsim$ 30--50 km \citep{2015SciA....115109J,2016IAUS..318....1K,2017ApJ...847L..12S,2017Sci...357.1026D,2018Icar..304...14T} are expected. All parameter models not listed in the \nameref{sec:methods} are identical to those used in \cite{2016Icar..274..350L} and \cite{Monteux2018}. In the planet formation model, the innermost disk radius is of the order 0.1 au and can vary over time. Disk lifetimes are distributed around 5 Myr, which is controlled via the photoevaporation rate \citep{2013A&A...549A..44F} and in agreement with current disk surveys \citep{2017ApJ...836...34M,Kral3}. The initial embryos are placed within the boundaries of [0.05, 40] au for G stars, and [0.086, 23.4] au for M stars. We vary in a Monte Carlo fashion \citep{2009A&A...501.1139M} the disk mass, lifetime, dust-to-gas ratio and the exponential cut-off radius \citep{2012A&A...547A.112M} to represent the diversity found in nature \citep{2010ApJ...723.1241A,2016ApJ...828...46A}.

\section*{Code and data availability}
The data that support the plots within this paper and other findings of this study are available from the corresponding author upon reasonable request.

\vspace{0.5cm}

\begin{acknowledgements}
The authors thank Richard Parker for helping to initiate this project, Edwin Kite, Thibaut Roger, Caroline Dorn for comments and discussions, and Martin Bizzarro and two anonymous referees for comments that helped to considerably improve the manuscript. T.L. was supported by ETH Z{\"u}rich Research Grant ETH-17 13-1 and acknowledges partial financial support from the Swiss Society for Astrophysics and Astronomy trough a MERAC travel grant. Y.A. acknowledges support from the Swiss National Science Foundation (SNSF). C.M. acknowledges support from the SNSF under grant BSSGI0$\_$155816 `PlanetsInTime'. The numerical simulations in this work were partially performed on the \textsc{EULER} computing cluster of ETH Z{\"u}rich. Parts of this work have been carried out within the framework of the National Center for Competence in Research PlanetS supported by the SNSF. We acknowledge the software usage of \textsc{matplotlib} \citep{Hunter:2007}, \textsc{scipy} \citep{scipy:2001}, \textsc{numpy} \citep{numpy:2011}, \textsc{pandas} \citep{pandas:2010}, and \textsc{seaborn} \citep{seaborn:2018}.
\end{acknowledgements}


\balance

\end{document}